\documentclass[aps,prb,twocolumn,showpacs,superscriptaddress,floatfix]{revtex4}
\usepackage{epsfig}
\usepackage{graphicx}
\usepackage{amsfonts}
\usepackage[figuresright]{rotating}
\usepackage{amssymb}
\usepackage{amsmath}
\usepackage{subfigure}

\def\avg#1{\langle#1\rangle}

\def\be{\begin{equation}} \def\ee{\end{equation}}
\def\bea{\begin{eqnarray}} \def\eea{\end{eqnarray}}

\def\nn{\nonumber}

\begin{document}
\title{Anisotropic vortex lattice structures in the FeSe superconductor}
\author{Hsiang-Hsuan Hung\footnote{On leave for
the Department of Electrical and Computer Engineering, University of
Illinois at Urbana-Champaign.} } \affiliation{Department of Physics,
University of California, San Diego, CA 92093}
\author{Can-Li Song}
\affiliation{State Key Laboratory for Low-Dimensional Quantum
Physics, Department of Physics, Tsinghua University, Beijing 100084,
China}
\author{Xi Chen}
\affiliation{State Key Laboratory for Low-Dimensional Quantum
Physics, Department of Physics, Tsinghua University, Beijing 100084,
China}
\author{Xucun Ma}
\affiliation{Institute of Physics, Chinese Academy of Sciences,
Beijing 100190, China
}
\author{Qi-kun Xue}
\affiliation{State Key Laboratory for Low-Dimensional Quantum
Physics, Department of Physics, Tsinghua University, Beijing 100084,
China}
\author{Congjun Wu}
\affiliation{Department of Physics, University of California, San Diego,
CA 92093}
\affiliation{Center for Quantum Information, IIIS, Tsinghua University, 
Beijing, China}

\begin{abstract}
In the recent work by Song {\it et al.} \cite{song2011}, the scanning
tunneling spectroscopy experiment in the stoichiometric FeSe reveals
evidence for nodal superconductivity and strong anisotropy.
The nodal structure can be explained with the extended $s$-wave pairing
structure with the mixture of the $s_{x^2+y^2}$ and $s_{x^2y^2}$ pairing
symmetries.
We calculate the anisotropic vortex structure by using the self-consistent
Bogoliubov-de Gennes mean-field theory.
In considering the absence of magnetic ordering in the FeSe at the ambient
pressure, orbital ordering is introduced, which breaks the $C_4$ lattice
symmetry down to $C_2$, to explain the anisotropy in the vortex tunneling
spectra.
\end{abstract}
\maketitle

%%%%%%%%%%%%%%%%%%%%%%%%%%%%%%%%%%%%%%%%%%%%%%%%%%%%%%%%%%%%%%%%%
\section{Introduction}
\label{sec:introcution}

Since the first iron-based layered superconductor La(O$_{1-x}$F$_x$)FeAs
had been discovered\cite{kamihara2008}, the family of iron-based
superconductors have brought huge impact in condensed matter physics community.
These novel materials exhibit similar phase diagrams compared to high-T$_c$
cuprates.
The parent compound LaOFeAs has an antiferromagnetic spin-density-wave
order \cite{lacruz2008}, and, upon doping, superconductivity appears.
Although correlation effects are weaker in iron-based superconductors
than those in high-T$_c$ cuprates, novel features arise from the
multi-orbital degree of freedom.
The orbital band structures play fundamental roles in determining
the Fermi surface configurations and pairing structures.

Understanding pairing symmetries is one of the most important issues
in the study of iron-based superconductors. Based on various
experimental \cite{hanaguri2010,ding2008,zhangxh2009} and
theoretical works
\cite{seo2008,daghofer2008,wang2008,tesanovic2009,tesanovic20092},
the nodeless $s_{x^2y^2}$-wave pairing has been proposed. In
momentum space, the Fermi surfaces of many iron-based
superconductors consist of hole pockets around the $\Gamma$-point,
and electron pockets around the two $M$-points. The signs of the
pairing order parameters on electron and hole Fermi surfaces are
opposite. The nodal lines of the gap function have no intersections
with Fermi surfaces, thus the $s_{x^2y^2}$-pairing is nodeless. In
the itinerant picture, the Fermi surface nesting between the hole
and the electron pockets facilitates the antiferromagnetic
fluctuations which favor the $s_{x^2y^2}$-pairing \cite{norman2008}.
In real space, an intuitive picture of the $s_{x^2y^2}$-wave pairing
is just the next-nearest-neighbor (NNN) spin-singlet pairing with
the $s$-wave symmetry \cite{seo2008}. Because the anion locations
are above or below the centers of the iron-iron plaquettes, the NNN
antiferromagnetic exchange $J_2$ is at the same order of the nearest
neighbor (NN) one $J_1$. The NNN $s_{x^2y^2}$-wave pairing can be
obtained from the decoupling of the $J_2$-term. On the other hand,
various experimental results have shown signatures of nodal pairing
structures \cite{fletcher2009,hicks2009,zeng2010}.
%One natural possibility is the $d_{x^2-y^2}$-pairing.
In the framework of the $s$-wave
pairing, the nodal pairing can be achieved through the
$s_{x^2+y^2}$-pairing \cite{mazin2008,yao2009,wang2010}. The
possibility for the $s_{x^2+y^2}$-wave pairing in iron-based
superconductors has also been shown in the functional
renormalization-group calculation \cite{wang2009}.

Another important aspect of the iron-based superconductors is the
spontaneous anisotropy of both the lattice and electronic degrees of
freedom, which reduces the 4-fold rotational symmetry to 2-fold. For
example, LaOFeAs undergoes a structural orthorhombic distortion and
a long-range spin-density wave (SDW) order at the wavevector $(\pi,0 )$
or $(0,\pi)$ \cite{lacruz2008}.
Similar phenomenon was also detected in NdFeAsO
by using polarized and unpolarized neutron-diffraction measurements
\cite{chen2008}. One popular explanation of the nematicity is the
coupling between lattice and the stripe-like SDW order \cite{fang2008,xu2008}.
The SDW ordering has also been observed in the FeTe system but with a 
different ordering wavevector at $(\frac{\pi}{2},\frac{\pi}{2})$
\cite{li2009,bao2009}.

Very recently, the experimental results of the FeSe superconductor
reported by Song {\it et al.}\cite{song2011} indicate a pronounced
nodal pairing structure in the scanning-tunneling-spectroscopy. Strong
electronic anisotropy is observed through the quasi-particle
interference of the tunneling spectra at much higher energy than the
superconducting gap. The low energy tunneling spectra around the
impurity and the vortex core also exhibit the anisotropy. The shapes
of the vortex cores are significantly distorted along one lattice
axis.

The anisotropy may arise from the structural transition from
tetragonal to orthorhombic phase at $90$K. However, the typical
orthorhombic lattice distortions in iron superconductors are at the
order of $0.012 \AA$, which is about half a percent of the lattice
constant and only lead to a tiny anisotropy in electronic
structures\cite{song2011}. In contrast, the anisotropic vortex cores
and impurity tunneling spectra observed by Song {\it et al.} are
clearly at the order of one. Therefore these anisotropies should be
mainly attributed to the electronic origin. The antiferromagnetic
long-range-order in such a system may be a possible reason for the
anisotropy.
For example, it has been theoretically investigated that the stripe-like 
SDW order can induce strong anisotropy in the quasi-particle interference 
of the STM tunnelling spectroscopy \cite{knolle2010}.
However, no evidence of magnetic ordering has been found
in FeSe at ambient pressure \cite{medvedev2009}, thus this
anisotropy should not be directly related to the long-range
magnetic ordering.

On the other hand, orbital ordering is another possibility for
nematicity in transition metal oxides.
For example, orbital ordering serves as a possible mechanism
for the nematic metamagnetic states observed in Sr$_3$Ru$_2$O$_7$
\cite{leewc2009,raghu2009}, and its detection through
quasi-particle interference has been investigated
\cite{leewc2010,leewc2009a}.
Orbital ordering has also been suggested to lift the degeneracy
between the $d_{xz}$ and $d_{yz}$-orbitals to explain the
anisotropy in iron-based superconductors.

In this article, we will study the effect of orbital ordering to the
vortex tunneling spectra in the FeSe superconductor. The rest of the
paper is organized as follows. In Sec. \ref{sec:hamiltonian}, a
two-band model Hamiltonian and the relevant band parameters are
introduced. The Bogoliubov-de Gennes mean-filed formalism is
described in Sec. \ref{sec:bdg-eq}. In Sec. \ref{sec:delatsymmetry},
we analyze the effects of orbital ordering to the tunneling spectra
of mixed pairing of the NN $s_{x^2+y^2}$-wave and the NNN
$s_{x^2y^2}$-wave in the homogeneous systems. In Sec.
\ref{sec:vortex}, the effects of orbital ordering to the anisotropic
vortex core tunneling spectra are investigated, which are in a good
agreement with experiments. 
Discussions and conclusions are given in Sec.
\ref{sec:summary}.

%%%%%%%%%%%%%%%%%%%%%%%%%%%%%%%%%%%%%%%%%%%%%%%%%%%%%%%%%%%%%%%%%
%%%%%%%%%%%%%%%%%%%%%%%%%%%%%%%%%%%%%%%%%%%%%%%%%%%%%%%%%%%%%%%%%
%%%%%%%%%%%%%%%%%%%%%%%%%%%%%%%%%%%%%%%%%%%%%%%%%%%%%%%%%%%%%%%%%

\section{Model Hamiltonian for the band structure}
\label{sec:hamiltonian}

\begin{figure}[!htb]
\centering\epsfig{file=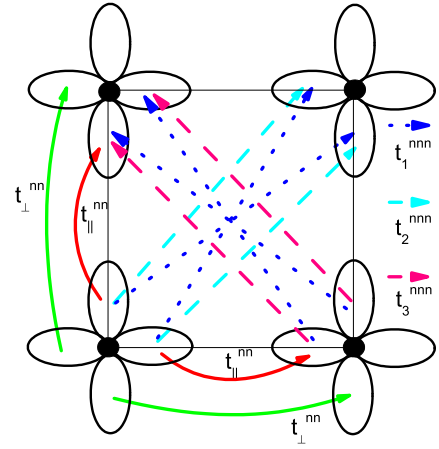,clip=1,width=0.65\linewidth,angle=0}
\caption{(Color online) The hopping schematic of the two-band tight-binding model
Eq. (\ref{eq:ham}) in a unit cell. Each black solid circle
represents an Fe atom. The solid arrows denote NN longitudinal
$\sigma$-bonding (red) and transverse $\pi$-bonding (green)
hoppings, respectively. The NNN intra-orbital hoppings $t^{nnn}_1$
are indicated by blue dot arrows along $\pm \hat{x}+\hat{y}$
directions. On the other hand, the NNN inter-orbital hoppings,
$t^{nnn}_2$ and $t^{nnn}_3$, along the $\hat{x}+\hat{y}$ and
$-\hat{x}+\hat{y}$ directions, respectively, are indicated by dash
arrows labeled with cyan and pink.} \label{fig:hopping}
\end{figure}

For simplicity, we use the two-band model involving the $d_{xz}$ and
$d_{yz}$-orbital bands in a square lattice with each lattice site
representing an iron atom, which was first proposed in Ref.
[\onlinecite{raghu2008}]. This is the minimal
model describing the iron-based superconductors, which can also
support orbital ordering. The tight-binding band Hamiltonian reads
\bea H_{0}&=& \sum_{\vec{r},\sigma}  \Big \lbrace
H^{nn}_{\parallel,\sigma} + H^{nn}_{\perp,\sigma}+
H^{nnn}_{\sigma}-\mu n_{\vec{r}} \Big \rbrace, \label{eq:ham} \eea
where \bea H^{nn}_{\parallel,\sigma}&=&t^{nn}_{\parallel} (
d^{\dag}_{xz,\sigma,\vec{r}} d_{xz,\sigma,\vec{r}+\hat{x}}+
d^{\dag}_{yz,\sigma,\vec{r}} d_{yz,\sigma,\vec{r}+\hat{y}} ), \nn \\
H^{nn}_{\perp,\sigma}&=& t^{nn}_{\perp} (
d^{\dag}_{xz,\sigma,\vec{r}} d_{xz,\sigma,\vec{r}+\hat{y}}+
d^{\dag}_{yz,\sigma,\vec{r}} d_{yz,\sigma,\vec{r}+\hat{x}} ), \nn \\
H^{nnn}_{\sigma}&=& t^{nnn}_1 (
 d^{\dag}_{xz,\sigma,\vec{r}} d_{xz,\sigma,\vec{r} \pm \hat{x}+\hat{y}}
+d^{\dag}_{yz,\sigma,\vec{r}} d_{yz,\sigma,\vec{r} \pm \hat{x}+\hat{y}} ) \nn \\
&+& t^{nnn}_2 (
 d^{\dag}_{xz,\sigma,\vec{r}} d_{yz,\sigma,\vec{r} + \hat{x}+\hat{y}}
 +d^{\dag}_{yz,\sigma,\vec{r}} d_{xz,\sigma,\vec{r}+ \hat{x}+\hat{y}} ) \nn \\
&+& t^{nnn}_3 (
 d^{\dag}_{xz,\sigma,\vec{r}} d_{yz,\sigma,\vec{r} - \hat{x}+\hat{y}}
+d^{\dag}_{yz,\sigma,\vec{r}} d_{xz,\sigma,\vec{r} - \hat{x}+\hat{y}}),
\nn \\
\eea where $d^{\dag}_{a,\sigma,\vec{r}}$ denotes the creation
operator for an electron with spin $\sigma$ on $d_{a}$-orbital
at site $\vec{r}$; $d_{a}$ refers to $d_{xz}$ and $d_{yz}$-orbitals;
$n_{\vec{r}}=n_{xz,\sigma,\vec{r}}+n_{yz,\sigma,\vec{r}}$
where $n_{a,\sigma,\vec{r}}=d^{\dag}_{a,\sigma,\vec{r}}d_{a,\sigma,\vec{r}}$
denotes the particle number operator; $\mu$ is the chemical
potential; $t^{nn}_{\parallel}$ and $t^{nn}_{\perp}$ denote the
longitudinal $\sigma$-bonding and transverse $\pi$-bonding between
nearest-neighboring (NN) sites, respectively; the three
next-nearest-neighboring (NNN) hoppings can be expressed in terms
of the NNN $\sigma$ and $\pi$-bondings $t^{nnn}_{\parallel}$ and
$t^{nnn}_{\perp}$, respectively as
$t^{nnn}_{1}=\frac{1}{2}(t^{nnn}_{\parallel}+t^{nnn}_{\perp})$,
$t^{nnn}_{2}=\frac{1}{2}(t^{nnn}_{\parallel}-t^{nnn}_{\perp})$, and
$t^{nnn}_{3}=\frac{1}{2}(-t^{nnn}_{\parallel}+t^{nnn}_{\perp})$.
We depict the hopping schematic of the two-band tight-binding
model Eq. (\ref{eq:ham}) in Fig. \ref{fig:hopping}.

%------------------------------------------------------------------
%---------------------------------------------------------------------

By introducing the spinor $\Psi(\vec{k})= [\psi_{xz,\sigma}(\vec{k}), 
\psi_{yz,\sigma}(\vec{k})]^T$ and
performing the Fourier transformation, the Hamiltonian in momentum
space becomes \bea H_{0}= \sum_{\vec{k}}
\Psi^{\dag}_{a,\sigma}(\vec{k}) \Big \lbrace H_{a
b}(\vec{k}) + \varepsilon(\vec{k}) - \mu \delta_{a b}
\Big \rbrace \Psi_{b,\sigma}(\vec{k}), \label{eq:kspace} \eea
where $a,b$ refer to the band index; the matrix kernel
$H_{a b}(\vec{k})$ is written as \bea \left (
\begin{array} {cc}
2t^{nn}_{\parallel}\cos{k_x}+2 t^{nn}_{\perp}\cos{k_y} &
4t^{nnn}_{3}\sin{k_x} \sin{k_y}\\
h.c. & 2t^{nn}_{\parallel} \cos{k_y}+ 2 t^{nn}_{\perp} \cos{k_x}
\end{array}  \right ), \nn
\eea and $\varepsilon(\vec{k})=2 \cos{k_x} \cos{k_y}
(t^{nnn}_{\parallel}+t^{nnn}_{\perp})$. This two-orbital model  has
also been used to study impurity resonance states
\cite{tsai2009,zhou2009}, vortex core states
\cite{hu2009,jiang2009}, and quasiparticle scattering
inference \cite{zhang2009}.

To fit the Fermi surface obtained from the LDA calculations
\cite{xu2008}, we use the parameter values below in the following
discussions as 
\bea t^{nn}_{\parallel}=0.8, ~t^{nn}_{\perp}=-1.4,
~t^{nnn}_{\perp}=1.8, ~ t^{nnn}_{\parallel}=0, 
\eea 
all which are in units of $t=1$, which is roughly at the 
energy scale around $100$meV.
This set of hopping integrals shows a
similar band structure, as the one Raghu {\it et al.} used. The
band-width is about $14$. In the following discussion, we use
$\mu=1.15$ which corresponds to slightly hole doped regimes. The
unfolded Brillouin zone (UBZ) embraces a hole surface around the
$\Gamma$ point [$\vec{k}=(0,0)$], and four hole pockets around the
$X$ point [$\vec{k}=(\pm \pi,\pm \pi)$] and four electron pockets
around $M$ point [$\vec{k}=(0,\pm \pi)$ or $(\pm \pi,0)$]
\cite{norman2008,lebegue2007}.

\begin{figure}[!htb]
\centering\epsfig{file=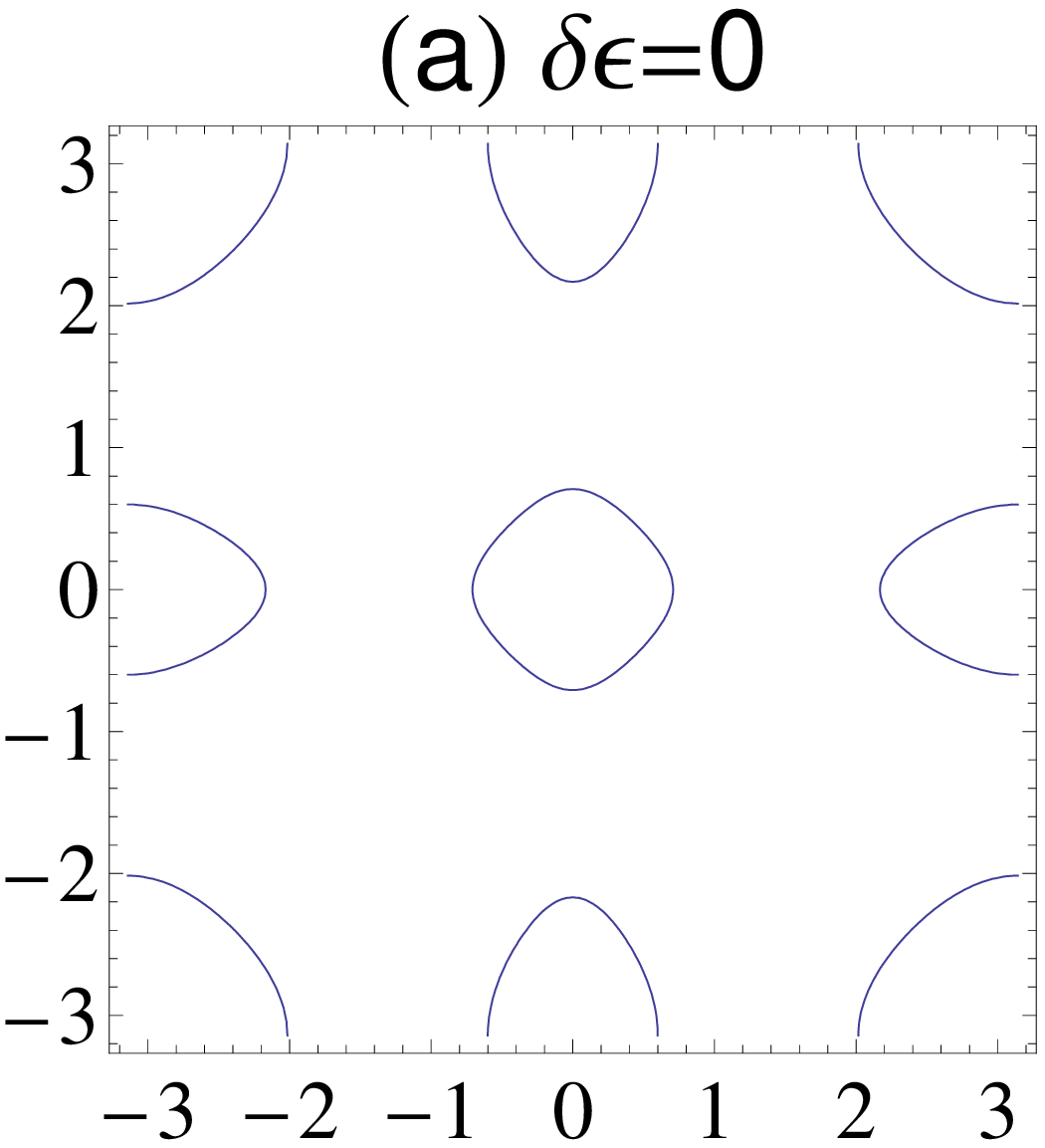,clip=0.7,width=0.45\linewidth,angle=0}
\centering\epsfig{file=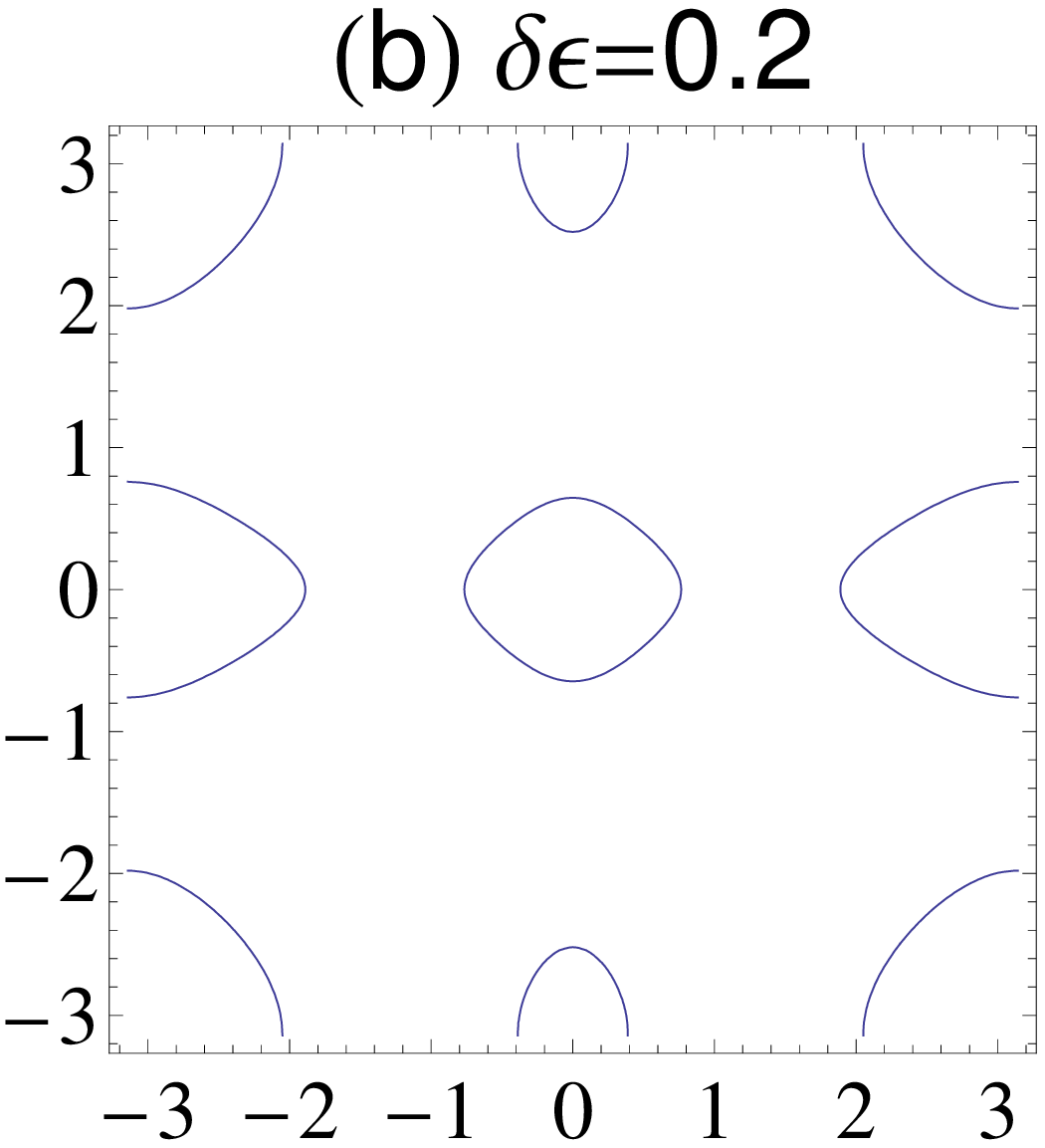,clip=0.7,width=0.45\linewidth,angle=0}
\caption{ Fermi surfaces in the UBZ for (a) $\delta \varepsilon=0$
and (b) $\delta\varepsilon=0.2$. The horizontal and vertical axes
denote $k_x$ and $k_y$, respectively. The parameter values used are
$t^{nn}_{\parallel}=0.8$, $t^{nn}_{\perp}=-1.4$,
$t^{nnn}_{\perp}=1.8$, $t^{nnn}_\parallel=0$, and $\mu=1.15$.
Anisotropic hole and electron pockets are shown in (b) at the
$\Gamma$ and $M$-points, respectively. } \label{fig:aniso}
\end{figure}

Our main purpose is to study the anisotropy effects in FeSe due to
the orbital ordering. Orbital ordering has been proposed in
iron-based superconductors in previous studies
\cite{kruger2009,lee2009,lv2009,chen2010,lv2010,lv2011}. Such an
ordering may arise from the interplay between orbital, lattice, and
magnetic degrees of freedom in iron superconductors. 
In this article, we are not interested in the microscopic mechanism of
spontaneous orbital ordering, but rather assume its existence to explain the
vortex tunneling spectra observed in Ref. [\onlinecite{song2011}].
According to the experimental data \cite{song2011}, strong anisotropy
has already been observed at least at the energy scale of 10meV, which is
much larger than the pairing gap value around 2meV.
Thus when studying superconductivity, we neglect the fluctuations
of the orbital ordering, but treat it as an external anisotropy.
For this purpose, we add an extra anisotropy term into the band structure
Eq. (\ref{eq:ham}) as
\bea 
H_{orb}= \delta
\varepsilon \sum_{\vec{r},\sigma} ( d^{\dag}_{xz,\sigma, \vec{r}}
d_{xz,\sigma, \vec{r}} - d^{\dag}_{yz,\sigma, \vec{r}} d_{yz,\sigma,
\vec{r}}), \label{eq:orbham} 
\eea 
which makes the $d_{xz}$-orbital energy higher than that of $d_{yz}$. 
For comparison, the Fermi
surfaces without and with the anisotropy term Eq. (\ref{eq:orbham})
are depicted in Fig. \ref{fig:aniso} (a) and (b), respectively. In
Fig. \ref{fig:aniso} (b), with the orbital term, the distortion of
the electron and hole pockets in $x$ and $y$-directions of the Fermi
surfaces appears such that the anisotropy is derived in the
iron-based superconductors.

%------------------------------------------------------------------
%---------------------------------------------------------------------

%%%%%%%%%%%%%%%%%%%%%%%%%%%%%%%%%%%%%%%%%%%%%%%%%%%%%%%%%%%%%%%%%
%%%%%%%%%%%%%%%%%%%%%%%%%%%%%%%%%%%%%%%%%%%%%%%%%%%%%%%%%%%%%%%%%
\section{The Bogoliubov-de Gennes formalism}
\label{sec:bdg-eq} In this section, we present the self-consistent
Bogoliubov-de Gennes (BdG) formalism based on the band structure
described in Sec. \ref{sec:hamiltonian}. In principle, for this
multi-orbital system, the general pairing structure should contain a
matrix structure involving both the intra and inter-orbital
pairings. Here for simplicity, we only keep the intra-orbital
pairing which is sufficient to describe the anisotropy observed in
the experiment by Song {\it et al.} \cite{song2011}.

The pairing interactions including the NN and NNN pairing are
defined as \bea \label{intham} H_{int}&=& - \frac{g_1}{2}
\sum_{\avg{ \vec{r},\vec{r}'}} \sum_{a}
\hat{\Delta}^{\dag}_{a}(\vec{r},\vec{r}')
\hat{\Delta}_{a}(\vec{r},\vec{r}'), \nn \\
&-& \frac{g_2}{2} \sum_{\avg{\avg{ \vec{r},\vec{r}'}}} \sum_{a}
\hat{\Delta}^{\dag}_{a}(\vec{r},\vec{r}')
\hat{\Delta}_{a}(\vec{r},\vec{r}'), \eea where $a$ is the
orbital index taking values of $d_{xz}$ and $d_{yz}$; $\langle
\vec{r}, \vec{r}' \rangle$ represents the NN-bonds and $\langle
\langle \vec{r}, \vec{r}' \rangle \rangle$ represents the NNN-bonds;
$g_{1,2}$ denotes the pairing interaction strengths along the NN and
NNN-bonds; $\hat{\Delta}_{a}(\vec{r},\vec{r}')$ describes the
spin singlet intra-orbital pairing operator across the bond defined
as \bea \label{pairingorder} \hat{\Delta}_{a}
(\vec{r},\vec{r}')=d_{a,\downarrow,\vec{r}}
d_{a,\uparrow,\vec{r}'}- d_{a,\uparrow,\vec{r}}
d_{a,\downarrow,\vec{r}'}. \eea where $\vec{r}'=\vec{r}+\vec
\delta$. For the $s_{x^2+y^2}$-pairing along the NN-bonds,
$\vec{\delta}=a \hat{x}(\hat{y})$, whereas for the
$s_{x^2y^2}$-pairing along the NNN-bonds, $\vec{\delta}=\pm a
(\hat{x}+\hat{y})$ NNN-bonds, where $a$ is Fe-Fe bond length,
defined as the lattice constant. In the square lattice, these two
pairings belong to the same symmetry class, thus they naturally
coexist. After the mean-field decomposition, the Hamiltonian becomes
\bea H_{MF}&=&H_0 - \frac{g_1}{2} \sum_{\langle \vec{r},\vec{r}'
\rangle} \sum_{a} \Delta^{*}_{a} (\vec{r},\vec{r}')
\hat{\Delta}_{a} (\vec{r},\vec{r}') \nn
\\ &-& \frac{g_2}{2} \sum_{\langle \langle \vec{r},\vec{r}' \rangle
\rangle} \sum_{a} \Delta^{*}_{a} (\vec{r},\vec{r}')
\hat{\Delta}_{a}(\vec{r},\vec{r}') + h.c., \label{eq:mixmfham}
\eea where $\Delta^{*}_{a}(\vec{r},\vec{r}')=\langle
\hat{\Delta}^{\dag}_{a}(\vec{r},\vec{r}') \rangle$ is the
pairing order parameter and  $\langle \cdots \rangle$ denotes the
expectation value over the ground state.

The mean-field BdG Hamiltonian Eq. (\ref{eq:mixmfham}) can be
diagonalized through the transformation as \bea \label{eq:bdgeq}
{c_{a,\uparrow}(\vec{r}) \choose c^{\dag}_{a,\downarrow}(\vec{r})}=
\sum_{n} \left (
\begin{array} {cc}
u_{a,n}(\vec{r}) & -v^{*}_{a,n}(\vec{r}) \\
v_{a,n}(\vec{r}) & u^{*}_{a,n}(\vec{r})
\end{array}  \right )
{\gamma_{a,n} \choose \gamma^{\dag}_{a,n}}. \eea The
eigenvectors associated with $E_n$ of the above BdG equations are
$(u_{a,n}(\vec{r}),v_{a,n}(\vec{r}))^T$ and the pairing
order parameters can be further obtained self-consistently as \bea
\Delta_{a}(\vec{r},\vec{r}')=\sum_n (u_n(\vec{r})
v^*_n(\vec{r}')+u_n(\vec{r}') v^*_n(\vec{r})) \tanh{\frac{b
E_n}{2}},
\nn \\
\eea where $b=1/k_BT$. After the wave functions are obtained
self-consistently, the local density of states (LDOS), which is
proportional to the conductance ($dI/dV$) in the scanning tunneling
microscopy, can be further measured by \bea \rho(\vec{r},E)&=&
\sum_{n,a} \Big\{ |u_{a,n}(\vec{r})|^2
L(E-E_n)\nn \\
&+&|v_{a,n}(\vec{r })|^2 L(E+E_n) \Big\}, \eea where $L(x)$ is
the Lorentzian as $L(x)=\gamma/[\pi(x^2+\gamma^2)]$ and $\gamma$
means the energy-broadening parameters, usually set around
$10^{-2}$.

When we study the vortex lattice structure problem, the single
particle Hamiltonian Eq. (\ref{eq:ham}) is modified by the magnetic
vector potential as 
\bea 
H_0^\prime=\sum_{\vec{r}_i,\vec{r}_j,
\sigma, a, b} t_{i,j; a, b} e^{i\frac{\pi}{\Phi_0}
\int^{\vec{r}_i}_{\vec{r}_j}\vec{A}(\vec{r}) \cdot d\vec{r}}
d^{\dag}_{a,\sigma, \vec{r}_i} d_{b, \sigma, \vec{r}_j},
\label{eq:megham} 
\eea where $\Phi_0=hc/2e$ is the quantized flux;
$a,b$ denote orbital indices and $\sigma$ denotes the spin index;
$t_{i,j;a,b}$ represents $t^{nn}_{\parallel}$, $t^{nn}_{\perp}$ or
$t^{nnn}_{1,2,3}$ depending on the corresponding bonds and orbitals.
The vector potential $\vec{A}(\vec{r})$ is chosen as the Landau
gauge by $(A_x,A_y)=(0,Bx)$.

Due to the magnetic translational symmetry, we can apply the
magnetic periodic boundary conditions to form Abrikosov vortex
lattices. Each magnetic unit cell carries a magnetic flux of
$2\Phi_0$, so that each magnetic unit cell contains two vortices. We
choose the size of the magnetic unit cell $pa \times qa$ with
$p=2q$, and the number of unit cells $N_x \times N_y$ with
$N_y=2N_x$. The corresponding magnetic field is $B=2 \Phi_0/pq
a^2=\Phi_0/(qa)^2$. In the Abrikosov vortex lattice, the translation
vector is written as $\vec{\textrm{V}}=(\textrm{X}pa,\textrm{Y}qa)$,
where $\textrm{X}=0,\cdots,N_x-1$ and $\textrm{Y}=0,\cdots,N_y-1$
are integers. The coordinate of an arbitrary lattice site can be
expressed as $\vec{R}=\vec{r}+\vec{\textrm{V}}$, where
$\vec{r}=(xa,ya)$ denotes the coordinate of the lattice site within
a magnetic unit cell, i.e. $1\le x \le p$ and $1 \le y \le q$.

Under the magnetic periodic boundary conditions, the eigenvectors of
the BdG equations Eq. (\ref{eq:bdgeq}) satisfy a periodic structure
written as \bea \left (
\begin{array} {c}
u_{a,n}(\vec{r}+pa\hat{x})  \\
v_{a,n}(\vec{r}+pa\hat{x})
\end{array}  \right ) &=&  e^{iK_x}\left (
\begin{array} {c}
e^{2\pi i \frac{y}{q}}u_{a,n}(\vec{r}) \\
e^{-2\pi i \frac{y}{q}}v_{a,n}(\vec{r})
\end{array}  \right ), \nn \\
\left (
\begin{array} {c}
u_{a,n}(\vec{r}+qa\hat{y})  \\
v_{a,n}(\vec{r}+qa\hat{y})
\end{array}  \right ) &=&  e^{iK_y}\left (
\begin{array} {c}
u_{a,n}(\vec{r}) \\
v_{a,n}(\vec{r})
\end{array}  \right ).
\eea Here $K_x=\frac{2\pi \textrm{X}}{N_x}$ and $K_y=\frac{2\pi
\textrm{Y}}{N_y}$ represent the magnetic Bloch wavevector on the x
and y components\cite{han2010}. With the relation, we can simulate
vortex lattices with sizes of $(N_xpa)\times (N_yqa)$ but reduce the
computational effort by diagonalizing $N_xN_y$ Hamiltonian matrices
with dimensions of $4pq$ rather than directly diagonalizing a
$4N_xN_ypq$ Hamiltonian matrix.

%%%%%%%%%%%%%%%%%%%%%%%%%%%%%%%%%%%%%%%%%%%%%%%%%%%%%%%%%%%%%%%%%
%%%%%%%%%%%%%%%%%%%%%%%%%%%%%%%%%%%%%%%%%%%%%%%%%%%%%%%%%%%%%%%%%
%%%%%%%%%%%%%%%%%%%%%%%%%%%%%%%%%%%%%%%%%%%%%%%%%%%%%%%%%%%%%%%%%

\section{The nodal {\it v.s.} nodeless pairings}
\label{sec:delatsymmetry}

In this section, we investigate the behavior of the superconducting
gaps in the homogeneous system. Due to the translation symmetry, the
pairing order parameters are spatially uniform, and we define
$\Delta_{a}(\vec \delta)=\Delta_{a} (\vec r, \vec r+\vec
\delta)$ where $a=d_{xz}, d_{yz}$. We start with the case in
the absence of orbital ordering, {\it i.e.}, $\delta \varepsilon=0$.
Due to the four-fold rotational symmetry, not all of the pairing
order parameters are independent. For the NN bond pairing, we have
$\Delta_{xz}(\hat x)= \Delta_{yz} (\hat y)$ and $\Delta_{xz} (\hat
y)=\Delta_{yz}(\hat x)$ due to the $s$-wave symmetry. As for the
NNN-bonding, the similar analysis yields the following relations of
$\Delta_{xz}(\hat{x}+\hat{y})= \Delta_{yz}(-\hat{x}+\hat{y})$, and
$\Delta_{xz}(\hat x -\hat y)=\Delta_{yz}(\hat x +\hat y)$.

Due to the multi-orbital structure, generally speaking, the pairing
order parameters have the matrix structure, thus the analysis of the
pairing symmetry is slightly complicated. However, before the
detailed calculation, we perform a simplified qualitative analysis
by considering the trace of the pairing matrix, defined as \bea
\Delta(\vec \delta)=\frac{1}{2} \big(\Delta_{xz}(\vec \delta)
+\Delta_{yz}(\vec \delta)\big) \label{eq:trace} \eea for both of the
NN and NNN-bonds. These quantities play the major role in
determining the pairing symmetry. The angular form factor of the
Fourier transform of the NN $s$-wave pairing is
$\Delta_{s_{x^2+y^2}}(k_x,k_y)\propto \cos{k_x}+\cos{k_y}$, and that
of the NNN $s$-wave is $\Delta_{s_{x^2y^2}}(k_x,k_y)\propto
\cos{k_x} \cos{k_y}$. 
Naturally $\Delta_{s_{x^2+y^2}}$ and $\Delta_{x^2y^2}$ have the same phase. 
Otherwise there would be a large energy cost corresponding to the 
phase twist in a small length scale of lattice constant.
The nodal lines of $\Delta_{s_{x^2+y^2}}$ have
intersections with the electron pockets, while the nodal lines of
$\Delta_{x^2y^2}$ have not intersections with Fermi surfaces.
Generally speaking, the NN and NNN $s$-wave pairings are mixed due
to the same symmetry representation to the lattice group as \bea
\Delta_{s_{\pm}}(k_x,k_y)=\Delta_1
(\cos{k_x}+\cos{k_y}) + \Delta_2 \cos{k_x} \cos{k_y}. \nn \\
\label{eq:mix-s-wave} \eea It is well-known that the gap function is
nodal for the NN $s$-wave pairing; while it is nodeless for the NNN
$s$-wave pairing. However, they can mix together. Recently this
aspect was supported by a variational Monte-Carlo
calculation,\cite{yang2010} where the authors discovered that the
$s_{x^2y^2}$-wave and $s_{x^2+y^2}$-wave states are energetically
comparable. Our BdG calculations show that even for the case of $g_2
\neq 0$ and $g_1 =0$, the $s$-wave NN pairing is still induced by
the $g_2$ term, and {\it vice versa} for the $s$-wave NNN pairing
showing that they can naturally coexist.

%-------------------------------------------------------------------------
\begin{figure}[!htb]
\centering\epsfig{file=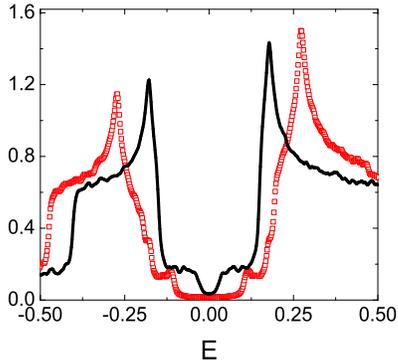,width=0.6\linewidth,
height=0.55\linewidth, angle=0} \caption{DOS {\it v.s.} tunneling
bias $E$ for the mixed $s$-wave pairing states at the zero
temperature. The red hollow squares with the parameter depict the
nodeless pairing with the dominant NNN pairing; and the black solid
line depicts the nodal pairing with the dominant NN pairing. The
parameter values are $(g_1=0.8, g_2=1.2)$ for the nodeless case and
$(g_1=1.2, g_2=0.4)$ for the nodal case, respectively. }
\label{fig:LDOS-swave}
\end{figure}
%--------------------------------------------------------------------------

%----------------------------------------------------------------
\begin{figure}[!htb]
\centering\epsfig{file=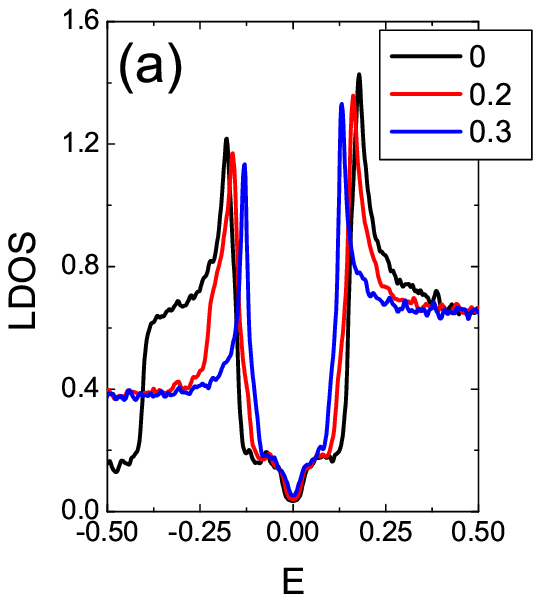,
width=0.48\linewidth,angle=0}
\centering\epsfig{file=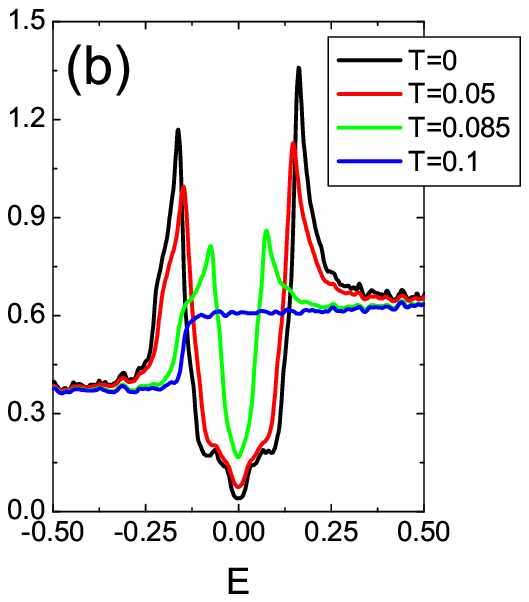,width=0.48\linewidth,angle=0}
\caption{ (a) DOS {\it v.s.} tunneling voltage $E$ for extended
$s_{\pm}$-wave without orbital anisotropy ($\delta \varepsilon=0$),
and the orbital anisotropy $(\delta \varepsilon =0.2, 0.3)$. Other
parameter values are $(g_1=1.2, g_2=0.4)$. At $\delta \varepsilon=0$
the coherent peaks are around $E=\pm 0.18$ and upon increasing
$\delta \varepsilon=0$ the locations of the coherent peaks only
slightly move towards zero energy. (b) The temperature dependence of
the DOS {\it v.s.} $E$ for the extended $s_{\pm}$-wave with orbital
anisotropy $\delta \varepsilon =0.2$ and the same parameters of
$g_{1,2}$ as in (a). At $T=0$, the coherent peaks are located at
$E=\pm 0.165$.} \label{fig:LDOS-OO}
\end{figure}
%---------------------------------------------------------------------

For the coexistence of the NN and NNN $s$-wave pairing, the pairing
gap function can be either nodal or nodeless. For the pure NNN
$s$-wave pairing, the nodal lines of the $\Delta_{x^2y^2}$ are
$k_{x,y}=\pm \frac{\pi}{2}$, which have no intersections with all of
the hole and electron pockets, thus the pairing is nodeless. For the
pure NN $s$-wave pairing, the nodal lines form a diamond box with
four vertices at the M-points $(\pm \pi, 0)$ and $(0, \pm \pi)$,
which intersects both electron pockets, thus the pairing is nodal.
When the NN and NNN $s$-wave pairings coexist, if the NN
$s_{x^2+y^2}$-wave pairing is dominant, the nodal lines of the
pairing function is illustrated in the Fig. S6 (a) of Ref.
[\onlinecite{song-supp}]. The original diamond nodal box is deformed
by pushing the four vertices way from the $M$-points to the
direction of the $\Gamma$-point. If the deformation is small, the
deformed diamond still intersects with the electron pockets, and
thus the pairing remains nodal. Upon increasing  the strength of the
NNN pairing, the deformation is enlarged and the intersections
disappear. Thus the pairing is nodeless.

Fig. \ref{fig:LDOS-swave}  reveals the density of states (DOS) {\it
v.s.} tunneling voltage for the mixed $s$-wave pairing state. The
red hollow squares are obtained using stronger NNN pairing strength
[{\it i.e.} $g_1=0.8$ and $g_2=1.2$ in Eq. (\ref{eq:mixmfham})],
providing gapful behavior which is similar to the pure $s_{x^2
y^2}$-wave state. In this case,
$\Delta_{xz}(\hat{x})=\Delta_{yz}(\hat{y}) = 0.068$,
$\Delta_{xz}(\hat{y})=\Delta_{yz}(\hat{x}) = -0.077$ and
$\Delta_{xz,yz}(\pm \hat{x} \pm \hat{y}) = 0.085$ showing stronger
NNN $s_{x^2y^2}$-wave pairing than NN $s_{x^2+y^2}$-wave pairing. On
the other hand, the black solid line (using $g_1=1.2$ and $g_2=0.4$)
indicates a gapless $V$-shape, similar to the pure
$s_{x^2+y^2}$-wave state. In contrary to the former case, the NN
pairing order parameters $\Delta_{xz}(\hat{x})=\Delta_{yz}(\hat{y})
= 0.071$, $\Delta_{xz}(\hat{y})=\Delta_{yz}(\hat{x}) = -0.059$ are
larger than NNN $s_{x^2y^2}$-wave ones $\Delta_{xz,yz}(\pm \hat{x}
\pm \hat{y}) = 0.057$. This reveals that the competition between the
gapful and gapless modes can be adjusted by tuning the ratios
between NNN and NN pairing interactions.

The main goal of our paper is the effect from orbital ordering which
is mimicked by Eq. (\ref{eq:orbham}). With the anisotropy, the
pairing structure  changes from Eq. (\ref{eq:mix-s-wave}) to the
following \bea \Delta_{s_{\pm}}(k_x,k_y)&=&\Delta_1
(\cos{k_x}+\lambda\cos{k_y}) \nn \\
&+& \Delta_2 \cos{k_x} \cos{k_y}, \eea where $\lambda$ is determined
by the anisotropy. The anisotropic nodal curve still intersects the
electron pockets as illustrated in the Fig. S6 (b) in Ref.
\onlinecite{song-supp}, and thus the superconducting gap functions
remain nodal. The calculated DOS vs energy patterns are plotted in
Fig. \ref{fig:LDOS-OO} (a) with the parameter values specified in
the figure caption. Upon finite $\delta \varepsilon$, the pairing
order parameters become anisotropic.
For example,  at $\delta \varepsilon=0.2$, we have
\bea
&&\Delta_{xz}(\hat{x})=0.059, \ \ \, \ \ \, \Delta_{yz}(\hat{y}) = 0.070, \nn \\
&&\Delta_{xz}(\hat{y})=-0.050, \ \ \, \Delta_{yz}(\hat{x})=-0.054, \nn \\
&& \Delta_{xz}(\hat{x}+\hat{y}) =
\Delta_{xz}(-\hat{x}+\hat{y})=0.0503 \nn \\
&&\Delta_{yz}(\hat{x}+\hat{y}) 
=\Delta_{yz}(-\hat{x}+\hat{y})=0.0505,
\eea 
and the gapless gap function and the $V$-shaped spectra remain in 
the moderate anisotropy from orbital ordering.

We also check the temperature dependence of DOS with the orbital
ordering as presented in Fig. \ref{fig:LDOS-OO} (b) with same
parameter values of $g_{1,2}$ as in Fig. \ref{fig:LDOS-OO} (a). With
orbital anisotropy $\delta \varepsilon=0.2$, the coherence peaks at
zero temperature is approximately $\Delta_{ch}\approx 0.165$. At low
finite temperatures, say $T=0.05\approx 0.3\Delta_{ch}$, the
$V$-shape DOS is still discernible. However, upon increasing
temperatures the $V$-shaped LDOS patterns smear  and eventually
the coherent peaks disappear at $T=0.1\approx 0.6 \Delta_{ch}$. In
this case the system turns into the normal state. This feature is
qualitatively consistent with the experimental observation of the
differential conductance spectra on FeSe in Ref.
[\onlinecite{song2011}].

%%%%%%%%%%%%%%%%%%%%%%%%%%%%%%%%%%%%%%%%%%%%%%%%%%%%%%%%%%%%%%%%%%%%%%%%%%%%
%%%%%%%%%%%%%%%%%%%%%%%%%%%%%%%%%%%%%%%%%%%%%%%%%%%%%%%%%%%%%%%%%%%%%%%%%%%%
\section{The vortex structure}
\label{sec:vortex}

In this section, we will study the vortex tunneling spectra for the
extended $s_{\pm}$-wave state. The size of the magnetic unit cell is
chosen as $pa \times qa = 20a \times 40a$, which contains two
vortices. The external magnetic field $B=2 \Phi_0/pqa^2$. The number
of magnetic unit-cells shown below is using $N_x\times N_y= 20
\times 10$, which is equivalent to the system size of $400a \times
400a$. The BdG equations are solved self-consistently with the
tight-binding model Eq. (\ref{eq:megham}) plus the mean-field
interaction  Eq. (\ref{intham}). The vortex configurations are
investigated for both cases with and without orbital ordering in
Sec. \ref{subsect:isotropy} and Sec. \ref{subsect:anisotropy},
respectively. The interaction parameters are $g_1=1.2$ and
$g_2=0.4$, and the temperature is fixed at zero.

%---------------------------------------------------------------
\subsection{Vortex structure in the absence of orbital ordering}
\label{subsect:isotropy}

\begin{figure}[!htb]
\epsfig{file=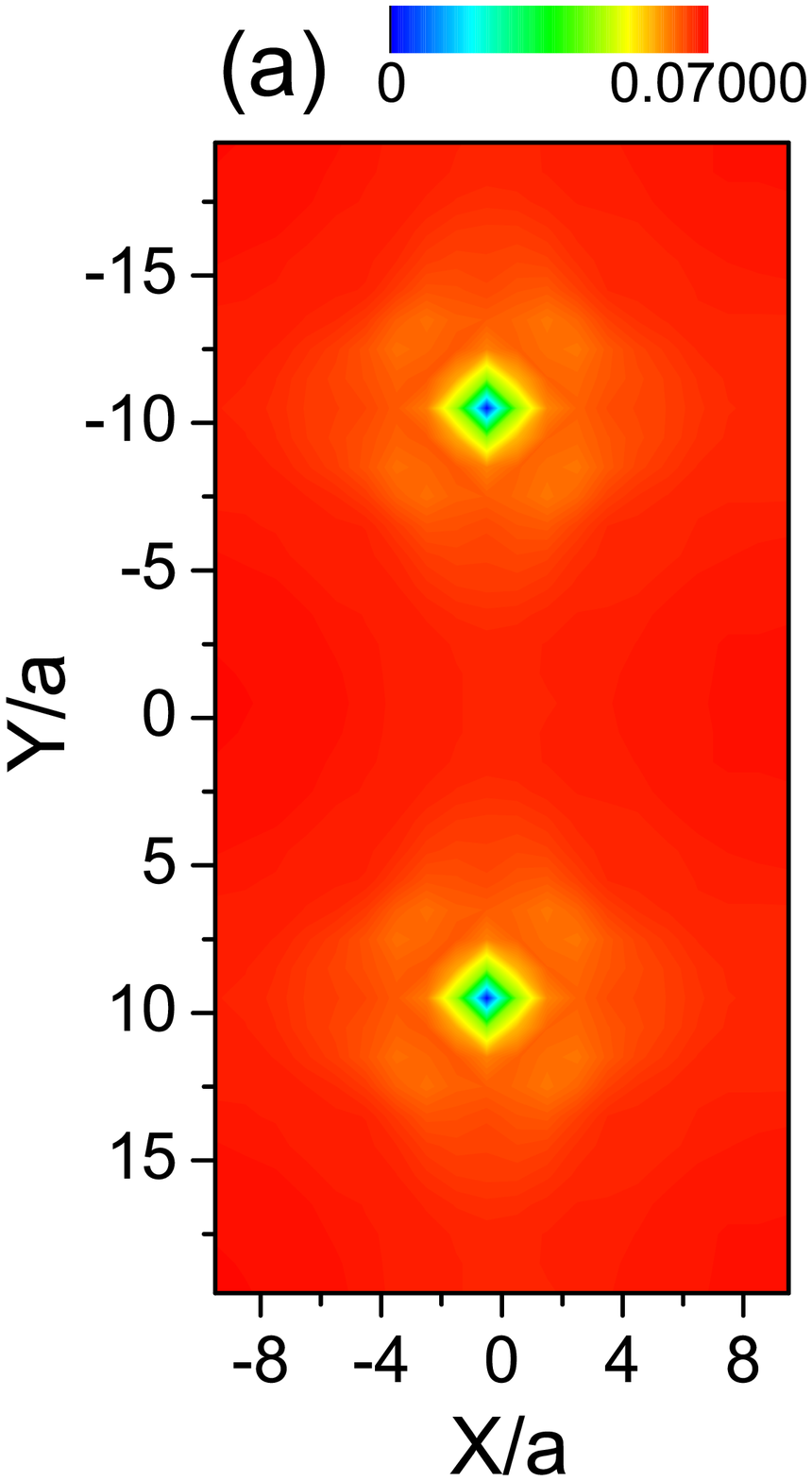,clip=0.7,width=0.35\linewidth,angle=0}
\epsfig{file=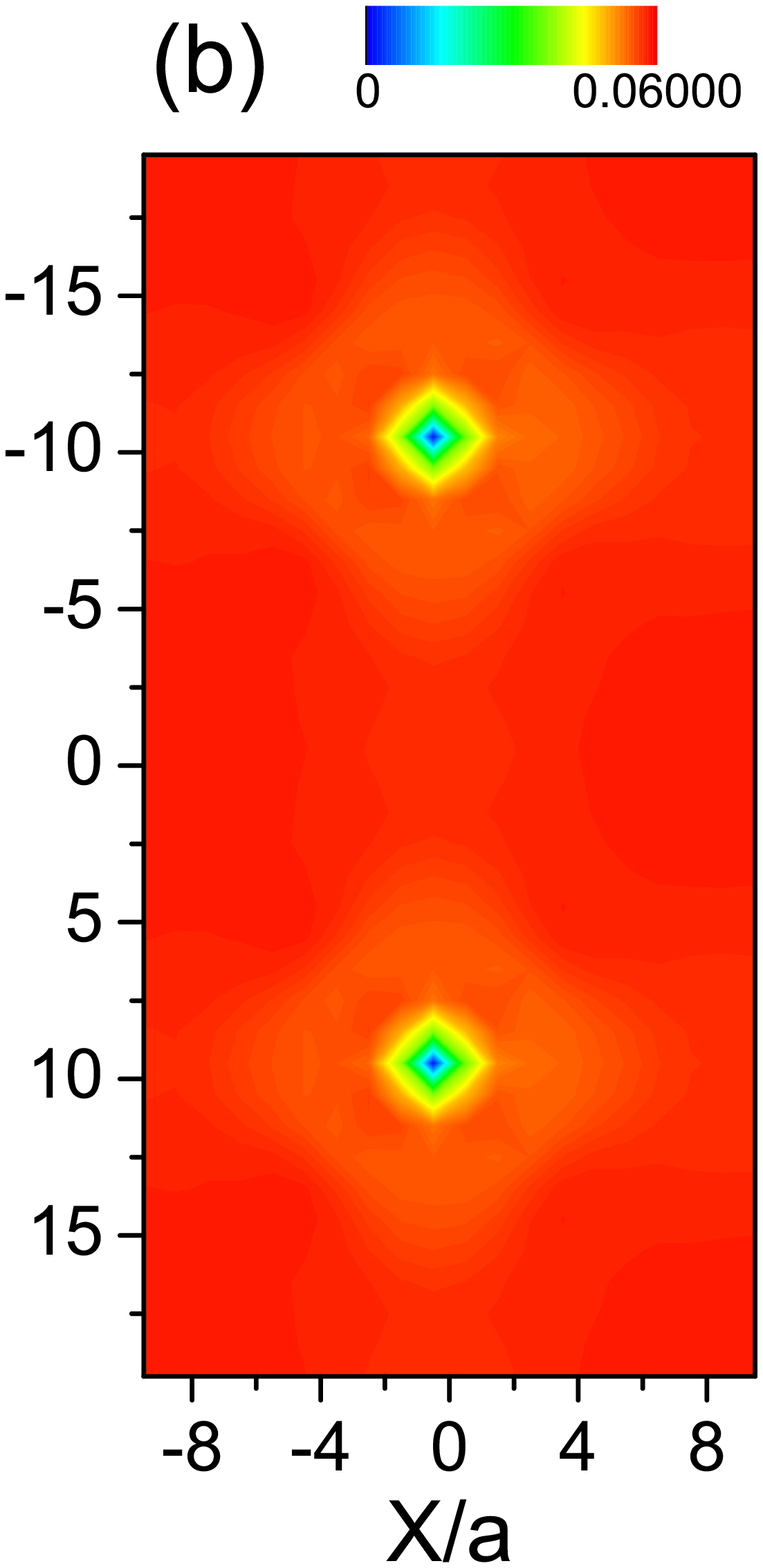,clip=0.7,width=0.312\linewidth,angle=0}
\epsfig{file=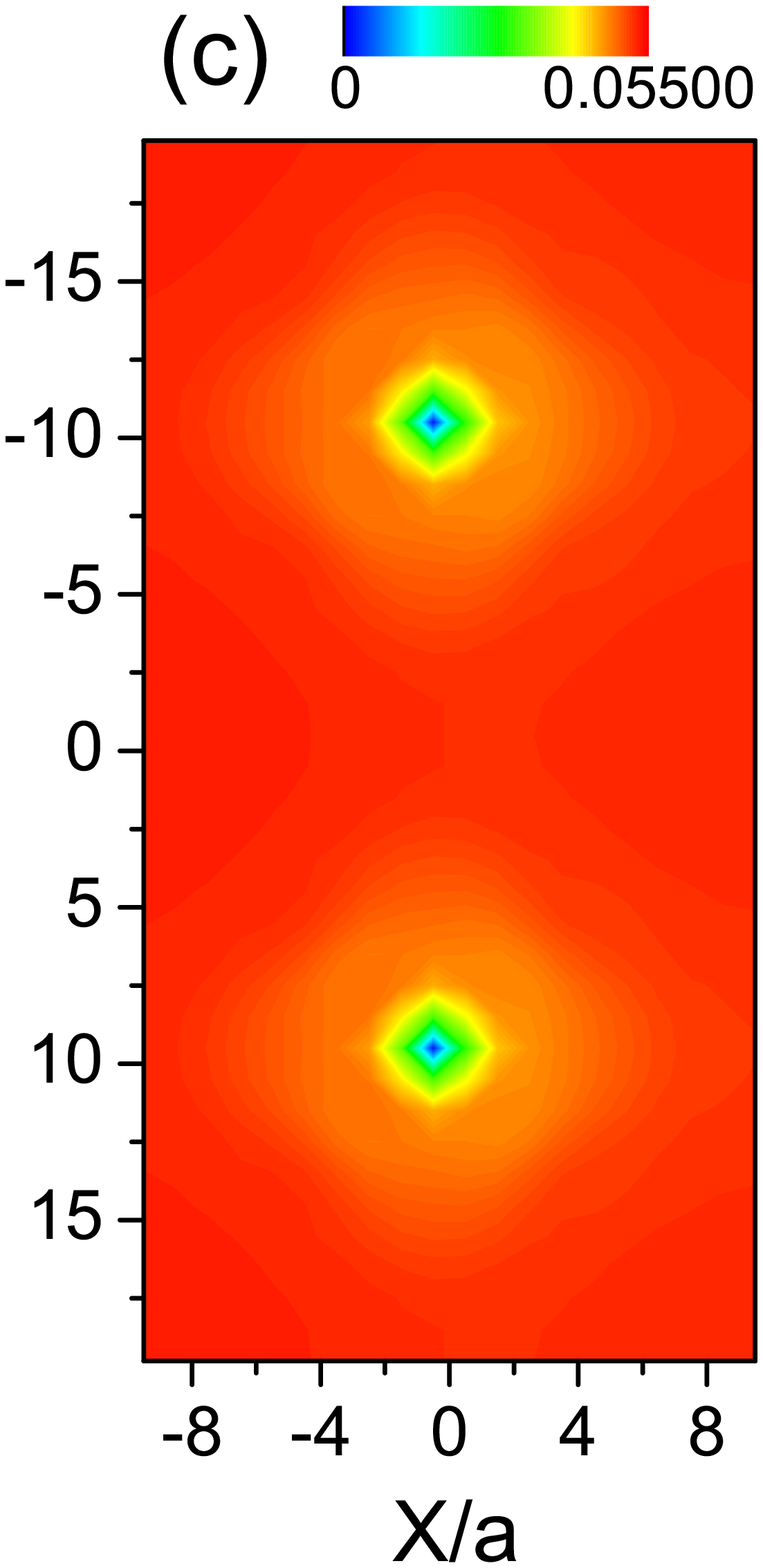,clip=0.7,width=0.31\linewidth,angle=0}\\
\epsfig{file=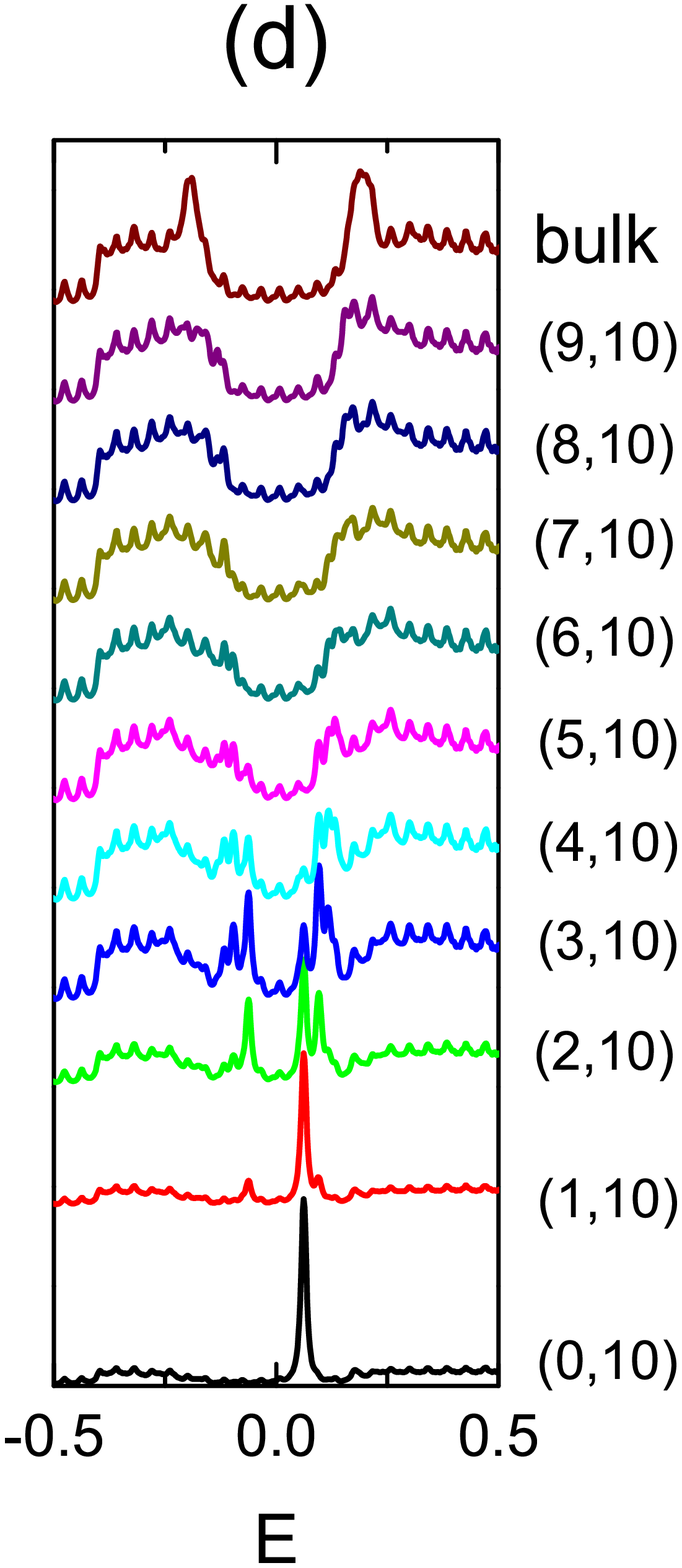,clip=0.7,width=0.4\linewidth,height=0.8
\linewidth, angle=0}
\epsfig{file=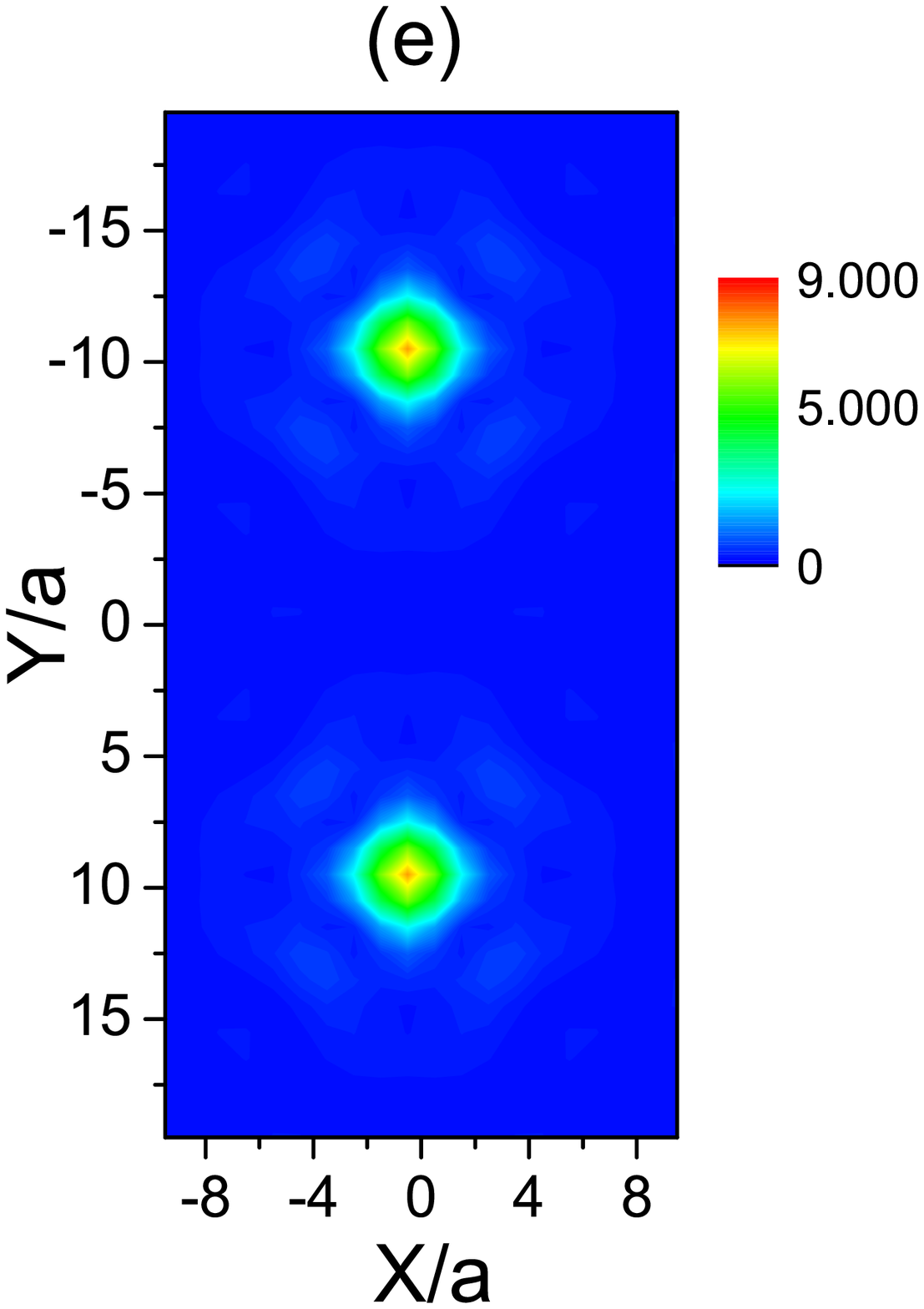,clip=0.7,width=0.55\linewidth,height=0.8
\linewidth,angle=0} \caption{The vortex structure for the mixed NN
$s_{x^2+y^2}$-wave NNN $s_{x^2y^2}$-wave pairing without orbital
ordering. (a) and (b) depict the spatial distribution of the
longitudinal and transverse NN pairings $\Delta^{NN}_{L}(\vec r)$
and $\Delta^{NN}_{T}(\vec r)$, respectively. (c) describes the NNN
$s$-wave pairing order parameters $\Delta^{NNN}(\vec{r})$. (d) The
LDOS {\it v.s.} {\it E} at different locations from the vortex core
[at $\vec{r}=(0,10a)$] to outside along the $x$-axis. The distances
of each site from the vortex core take the step of one lattice
constant. (e) The spatial LDOS distribution at the energy of the
vortex core resonance peak $E_{re}\approx 0.062$.}
\label{fig:sw-plain-delta}
\end{figure}
%-------------------------------------------------------------------

We start with the vortex configuration without orbital ordering. The
NN $s_{x^2+y^2}$-wave and NNN $s_{x^2y^2}$-wave pairing order
parameters in real space as defined as follows. We define the
longitudinal and transverse NN $s$-wave pairings as \bea
\Delta^{NN}_{L}(\vec{r})&=&\frac{1}{4} \Big\{ \Delta_{xz}(\vec r,
\vec r + a\hat x)+
\Delta_{xz}(\vec r, \vec r - a\hat x)\nn \\
&+&
\Delta_{yz}(\vec r, \vec r + a\hat y)+
\Delta_{yz}(\vec r, \vec r - a\hat y)\Big\}\nn \\
\Delta^{NN}_{T}(\vec{r})&=&\frac{1}{4}
\Big\{\Delta_{yz}(\vec r, \vec r + a\hat x)+
\Delta_{yz}(\vec r, \vec r - a\hat x)\nn \\
&+&
\Delta_{xz}(\vec r, \vec r + a\hat y)+
\Delta_{xz}(\vec r, \vec r - a\hat y)\Big\}.
\eea
For the NNN pairing related to site $\vec r$, we define
\bea
\Delta^{NNN} (\vec{r})=\frac{1}{8}
\sum_{a=xz,yz;\vec{\delta}=\pm \hat{x} \pm \hat{y}} \Delta_{a}
(\vec{r},\vec{r}+\vec{\delta}).
\label{eq:swavepair}
\eea
The pure NN $s_{x^2+y^2}$-wave vortex states are recently investigated to
describe the competition between the superconductivity and
spin-density-wave (SDW) in the hole-doped materials,
Ba$_{1-x}$K$_x$Fe$_2$As$_2$ \cite{gao2010}. Similar calculation for
the pure NNN $s_{x^2 y^2}$-wave with SDW has been used to study
BaFe$_{1-x}$Co$_x$As$_2$ \cite{hu2009} and the FeAs stoichiometric
compounds\cite{jiang2009}. In our case, these two $s$-wave pairing
order parameters mix together.

The real space profiles of the longitudinal and transverse NN
$s$-wave pairings $\Delta^{NN}_L(\vec r)$ and $\Delta^{NN}_T(\vec
r)$ are depicted in Fig. \ref{fig:sw-plain-delta} (a) and (b),
respectively. The vortex cores are located at $\vec{r}=(0,\pm 10 a)$
where the pairing order parameters are suppressed. Both of them
exhibit the $C_4$ symmetry. The NNN $s$-wave pairing order
parameters are depicted in Fig. \ref{fig:sw-plain-delta} (c) and the
vortices are diamond-shaped with the $C_{4}$ rotational symmetry.
Note that the maximum magnitudes of the NNN $s_{x^2y^2}$-wave
pairing order parameters are smaller than those of NN longitudinal
and transverse $s_{x^2+y^2}$-wave pairing order parameters, due to
the stronger NN pairing strength ($g_1=1.2$ and $g_2=0.4$). The
coherence length can be estimated as $\xi \approx 4a \sim 5a$ from
the spatial distributions of order parameters.

The relations of LDOS {\it v.s.} the tunneling energy $E$ are
presented in Fig. \ref{fig:sw-plain-delta} (d) at different
locations from the vortex core to outside along $x$-axis. The LDOS
pattern along $y$-axis is the same as Fig. \ref{fig:sw-plain-delta}
(d) due to $C_4$ symmetry. Note that there exist fine
oscillations in the LDOS pattern. This fine oscillation structure
may come from the Landau oscillation in which the oscillation period
is related to the external magnetic field\cite{hu2009}.
At the
vortex core [$\vec{r}=(0,10a)$], the coherence peaks at the bulk gap
value of $\Delta_{ch}$ disappear.
Instead, a resonance peak appears at $E_{re}\approx 0.062$.  Away from the
vortex core, the resonance peak splits into the particle and hole
branches with energies which symmetrically distribute with respect
to $E_f$. As the distance increases,  the peak intensities decrease,
and the energy separations between the particle and hole branches of
peaks increase. As the distance reaches around $6a$, i.e. beyond the
coherence length $\xi$, these peaks merge into the bulk coherence
peaks.
Fig. \ref{fig:sw-plain-delta} (e) presents the spatial
distribution of the LDOS at the vortex core resonance state energy
$E_{re}$, which exhibits the 4-fold rotational symmetry. The vortex
core state mainly distributes within one coherence length, thus it
is closer to a bound state rather than a resonance state.

%---------------------------------------------------------
\subsection{The vortex structures with orbital ordering}
\label{subsect:anisotropy}

%------------------------------------------------------------
\begin{figure}[!htb]
\epsfig{file=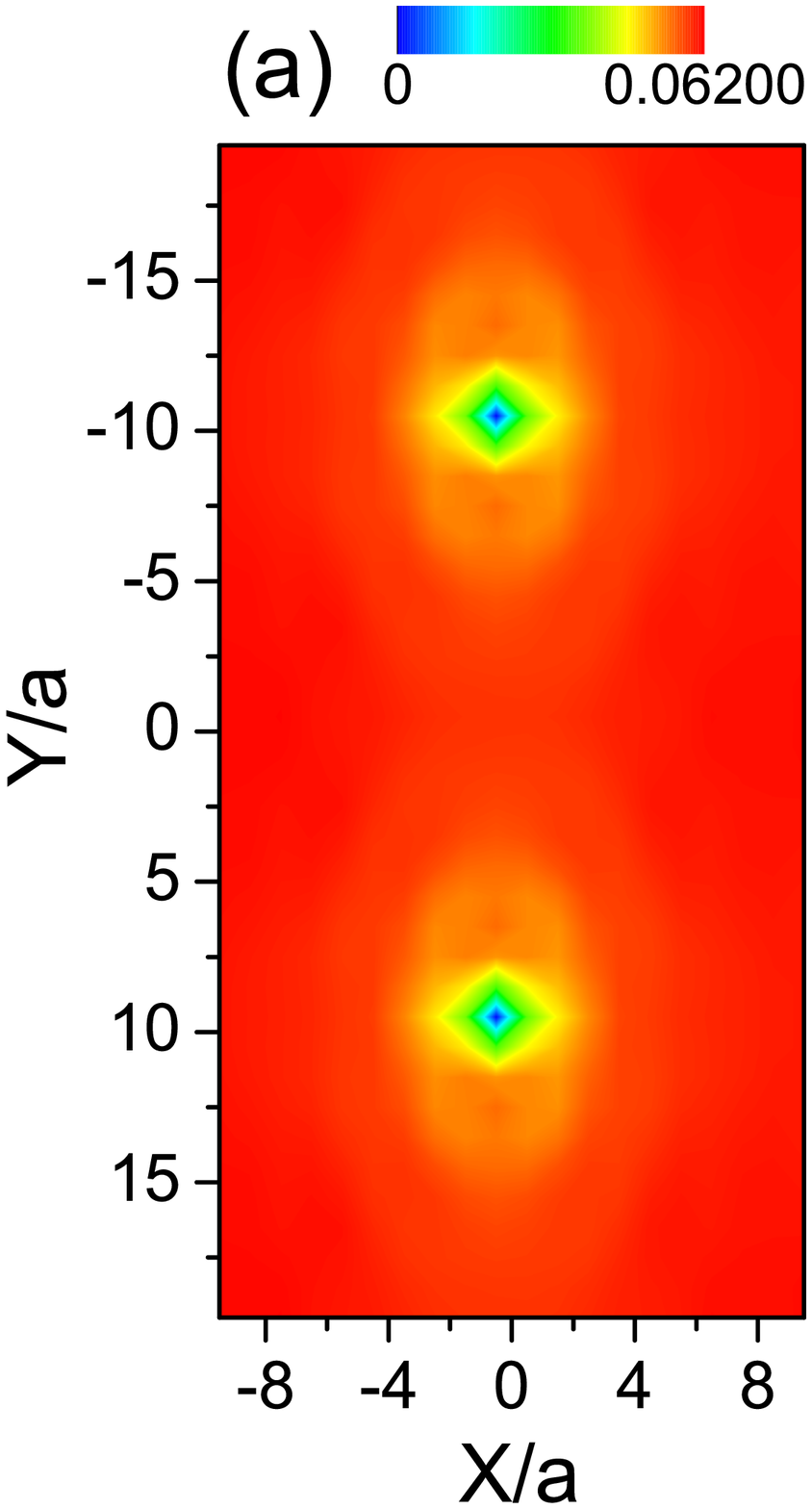,clip=0.7,width=0.341\linewidth,angle=0}
\epsfig{file=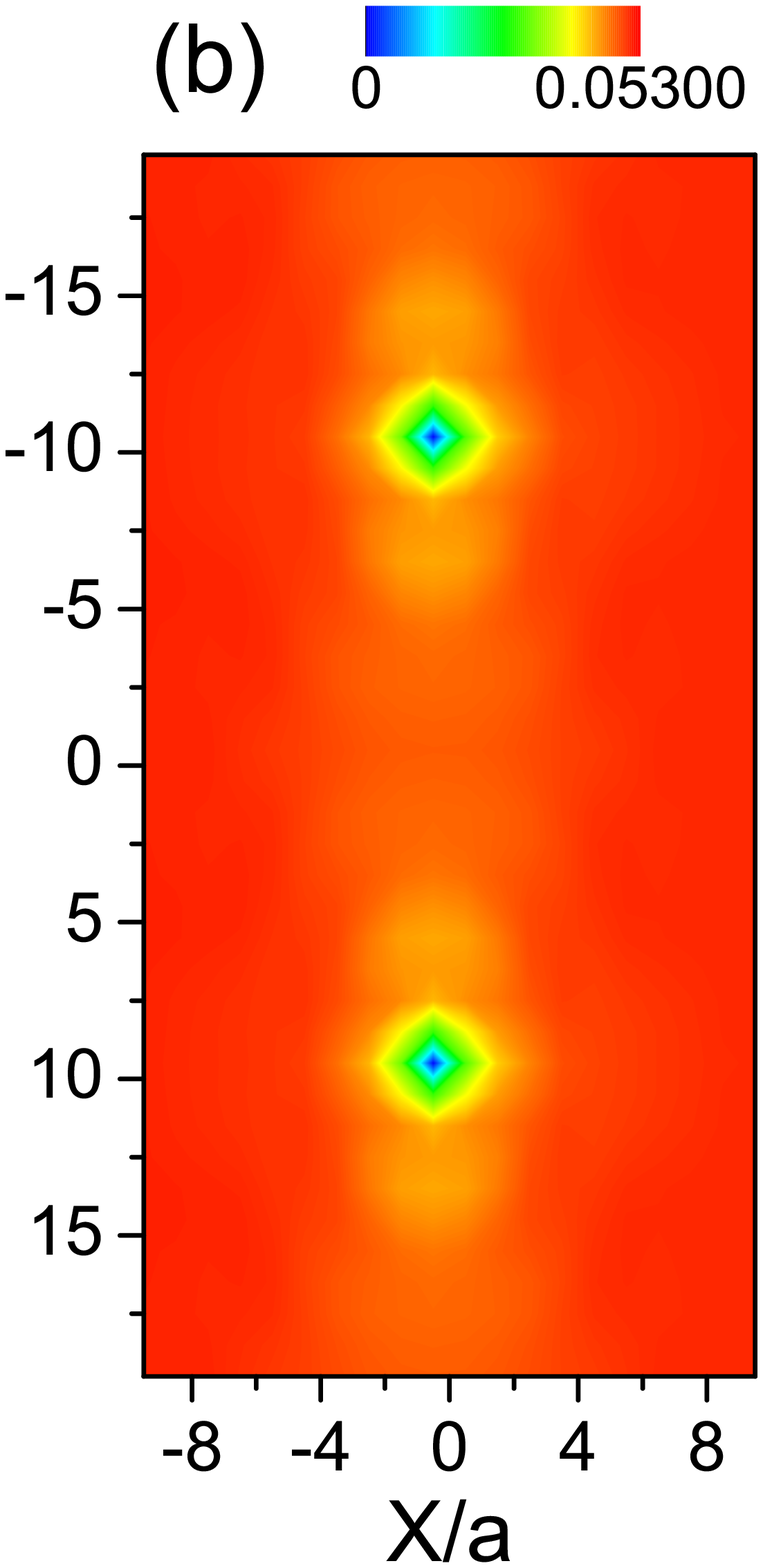,clip=0.7,width=0.31\linewidth,angle=0}
\epsfig{file=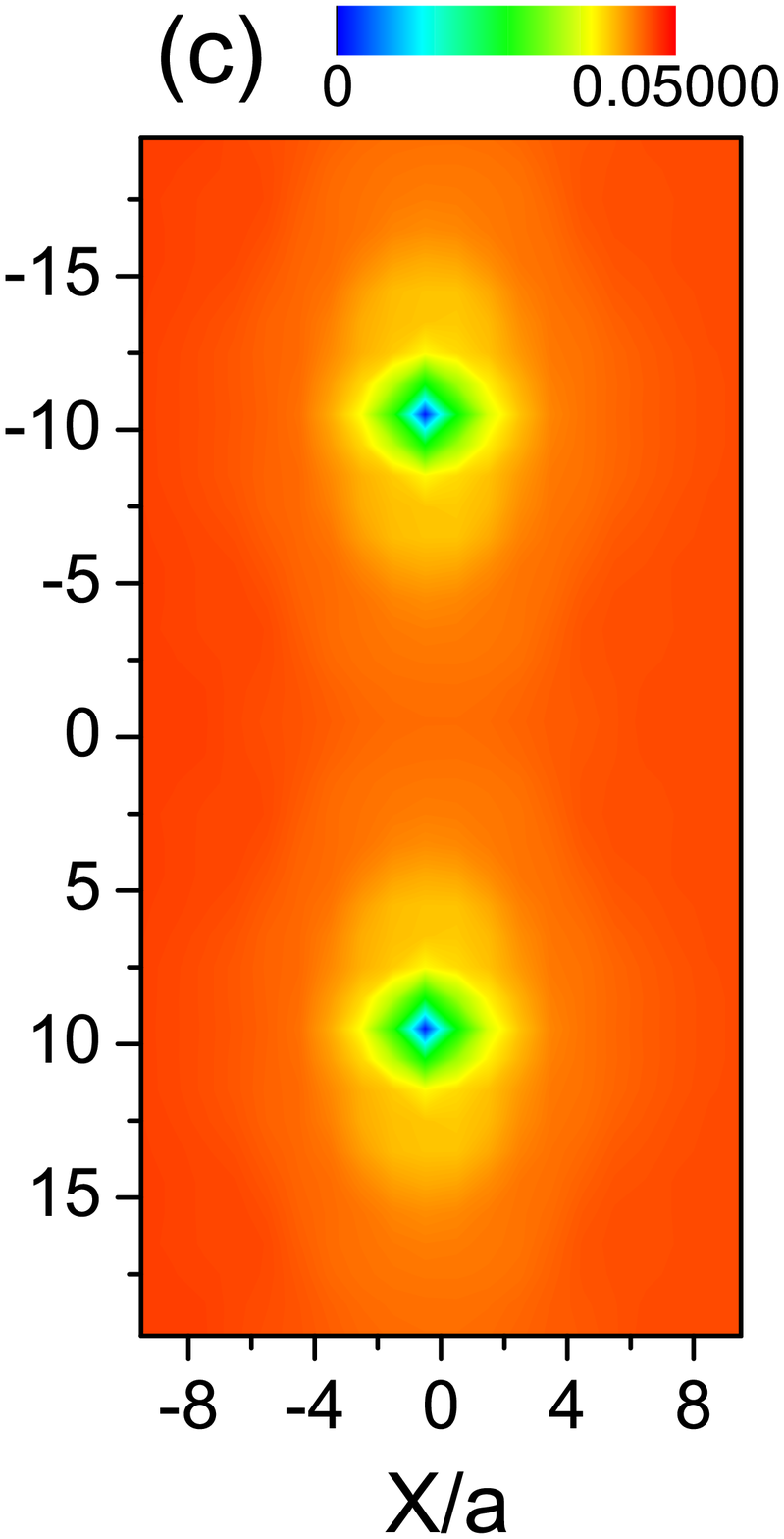,clip=0.7,width=0.322\linewidth,angle=0}
\caption{The spatial distributions of the $s$-wave pairing order
parameters with the orbital anisotropy $\varepsilon=0.2$. (a) The
longitudinal NN pairing $\Delta^{NN}_{L}(\vec r)$, (b) the
transverse NN pairing $\Delta^{NN}_{T}(\vec r)$, and (c) the NNN
$s_{x^2y^2}$-wave pairing order parameters $\Delta^{NNN}(\vec r)$.
All of them show anisotropy.} \label{fig:sw-OO-delta}
\end{figure}
%------------------------------------------------------------

In this subsection, we consider the effect of orbital ordering to
the vortex lattice states.
The band and interaction parameters are the same as in
Sec. \ref{subsect:isotropy}, except that we  add the anisotropy term of
Eq. (\ref{eq:orbham}) with $\delta \varepsilon=0.2$. Such a term
breaks the degeneracy between the $d_{xz}$ and $d_{yz}$-orbitals,
and reduces the $C_{4}$ symmetry down to $C_2$.

Fig. \ref{fig:sw-OO-delta} (a), (b) and (c) depict the spatial
distributions of the NN $s_{x^2+y^2}$-pairing order parameters
$\Delta^{NN}_{L}$ and $\Delta^{NN}_{T}$, and the NNN
$s_{x^2y^2}$-pairing order parameters in a magnetic unit cell,
respectively. All of them clearly exhibit the breaking of the 4-fold
symmetry down to the 2-fold one. For the dominant NN
$s_{x^2+y^2}$-pairings, the coherence lengths along the $x$ and
$y$-directions are no longer the same, which can be estimated as
$\xi_x\approx 3a$ and $\xi_y \approx 7a$, respectively. From Fermi
surface Fig. \ref{fig:aniso} (b), in the presence of orbital
ordering, the electron pocket in the $y$-direction shrinks, which
implies that Cooper pairing superfluid stiffness is weaker along the
$y$-direction than the $x$-direction. This picture agrees with the
larger value of $\xi_y$ exhibiting in Figs. \ref{fig:sw-OO-delta}.

%---------------------------------------------------------------
\begin{figure}[!htb]
\epsfig{file=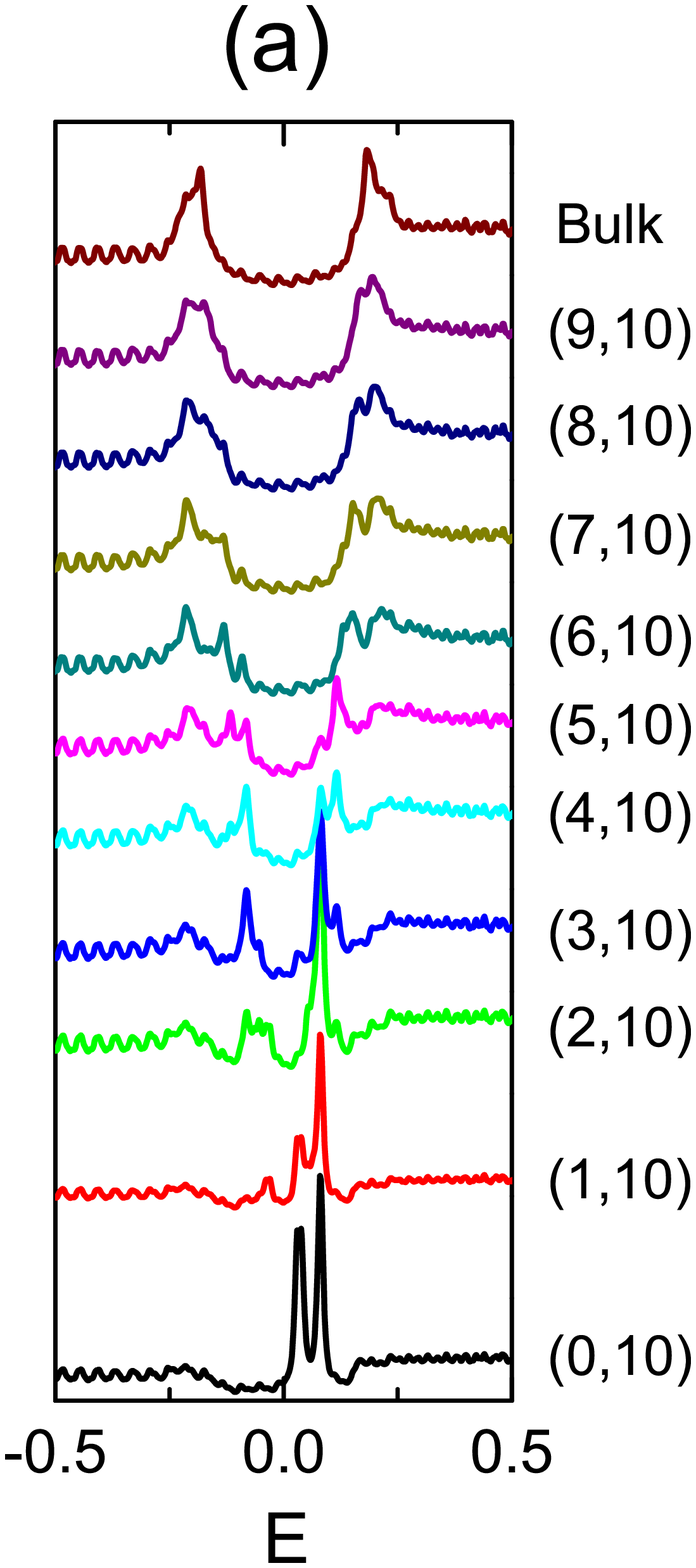,clip=0.7,width=0.453\linewidth,angle=0}
\epsfig{file=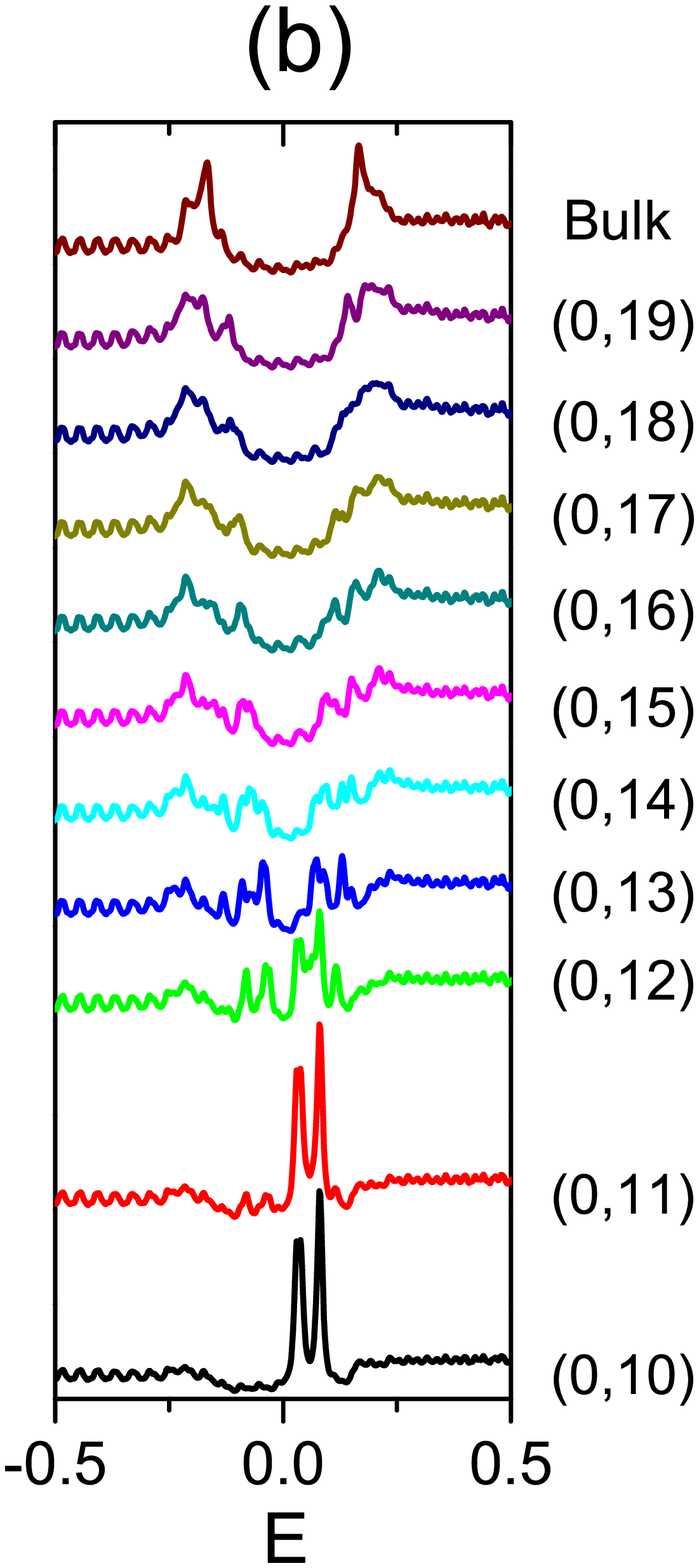,clip=0.7,width=0.455\linewidth,angle=0}
\\
\epsfig{file=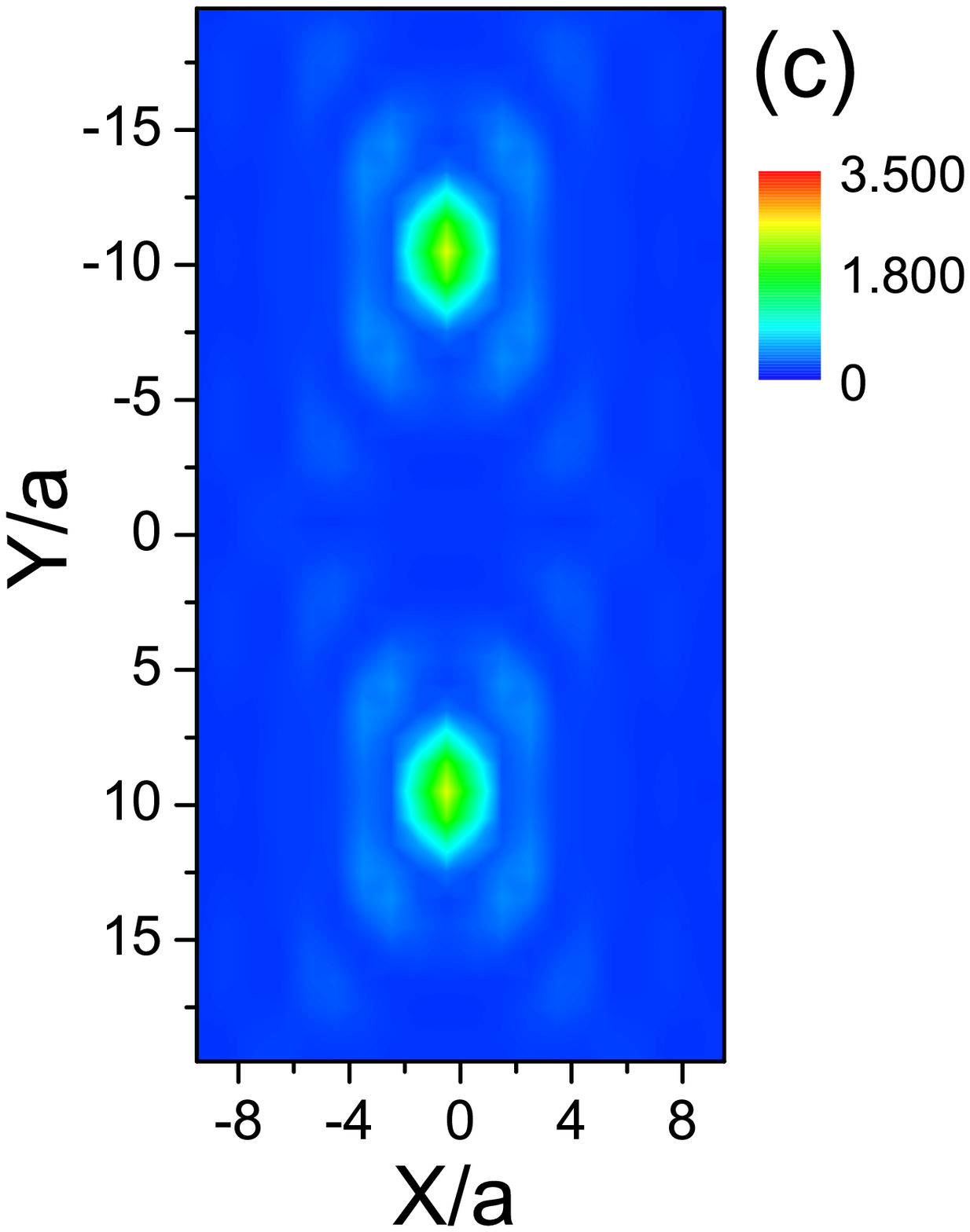,clip=0.7,width=0.6\linewidth,angle=0}
\caption{The LDOS {\it v.s.} tunneling energy $E$ at different sites
from vortex core to outside along (a) the $x$-axis (short axis) and
(b) the $y$-axis (long axis). The distances of each site from the
vortex core take the step of one lattice constant. (c) The spatial
LDOS distributions at the vortex core resonant energy $E_{er}=0.038$
.} \label{fig:sw-OO-LDOS}
\end{figure}
%------------------------------------------------------------

Next we turn to study the LDOS patterns for the extended
$s_{\pm}$-wave with orbital ordering. In comparison with those
without orbital ordering depicted in Fig. \ref{fig:sw-plain-delta}
(c), the LDOS patterns in Fig. \ref{fig:sw-OO-LDOS} (a) and (b)
exhibit significant anisotropy.
At the vortex core center, the resonance peak splits into two pieces
inside gap.
This feature is similar to the previous studies on the
$s_{x^2 y^2}$-wave pairing with SDW in iron-based
superconductor\cite{hu2009} and the $d$-wave pairing with
antiferromagnetic ordering in cuprates\cite{zhu2001}. Away from the
vortex center, however, the peaks of the particle and hole branches
along short ($x$) and long ($y$) axes behave differently. The
separation between the particle and hole peaks disperses along the
short axis much quicker than that along the long axis. The peak
intensities along the short axis are stronger than those along the
long axis. These features are in good agreement with recent
experiment observations \cite{song2011}.
The pronounced anisotropy also exhibits in the spatial variation of
the LDOS at the energy $E_{er}=0.038$ where the resonance peak is located,
presented in Fig. \ref{fig:sw-OO-LDOS} (c).
Moreover, with negative $\delta
\varepsilon$ which loads particles in $d_{xz}$ in prior to $d_{yz}$,
all above pictures will have a $\pi /2$-rotation. Therefore, the
experimentally observed anisotropy agrees with the picture of
orbital ordering.

%%%%%%%%%%%%%%%%%%%%%%%%%%%%%%%%%%%%%%%%%%%%%%%%%%%%%%%%%%%%
\section{Discussions and Conclusions}
\label{sec:summary}

We have studied a minimal two-orbital model with orbital ordering to
interpret the recent STM observations in Ref.
[\onlinecite{song2011}] including the nodal superconductivity and
the anisotropy vortex structure in FeSe superconductors. 
In considering the absence of magnetic long-range-order in FeSe at ambient 
pressure, orbital ordering provides a natural formalism for anisotropy. 
The NN $s_{x^+y^2}$-wave and the NNN $s_{x^2y^2}$-wave are considered, which 
generally are mixed. 
When the NNN pairing dominates, nodal pairing still exists even in the 
appearance of orbital ordering. We further performed the BdG calculation for the
vortex tunneling spectra in the presence of orbital ordering, which
breaks $C_4$ rotational symmetry down to $C_2$ and explicitly
induces anisotropic vortex structures.

The microscopic mechanism of the origin of this spontaneous anisotropy 
remains an open question. 
It might be related to the strong antiferromagnetic fluctuations.
As shown in Ref. [\onlinecite{fernandes2012,xu2008,fang2008}], 
before the onset of the antiferromagnetic long rang order, the
spin nematic order with a Z$_2$ symmetry breaking occurs.
This nematic order corresponds to anisotropic spin-spin correlation 
along $a$ and $b$-axis, which can in turn induce orbital ordering.

\section{Acknowledgement}
C. W. thanks Z. Y. Lu, and F. J. Ma for helpful discussion.
H. H. H. and C. W. are partly supported by the NBRPC (973 program) 
2011CBA00300 (2011CBA00302),  NSF-DMR-1105945 and Sloan
Research Foundation. C. L. S., X. C., X. C. M. and Q. K. X. are
supported by National Science Foundation and Ministry of Science and
Technology of China.

{\it Note added} \ \ \ Near the completion of this manuscript, we
learned the article by Chowdhury {\it et al.} \cite{chowdhury2011}
which studied the anisotropic vortex tunneling spectra in Ref.
[\onlinecite{song2011}] through the Ginzburg-Landau formalism.

\end{document}